\documentclass[12pt,oneside]{article}
\usepackage{graphicx}
\def\be{\begin{equation}}
\def\ee{\end{equation}}
\def\ba{\begin{array}}
\def\ea{\end{array}}
\def\bea{\begin{eqnarray}}
\def\eea{\end{eqnarray}}
\def\<{\langle}
\def\>{\rangle}
\def\~{\tilde}
\def\s{\sigma}

\def\t{\tau}

\newtheorem{theorem}{Theorem}

\newtheorem{definition}[theorem]{Definition}

\newcommand{\brac}[1]{\< #1\>}

\newcommand{\SN}{\Sigma_N}
\newcommand{\mod}{\,\,\mbox{mod }}

\begin{document}
\title{Energy Landscape Statistics of the \\
Random Orthogonal Model} \maketitle
\begin{center}
\author{M. Degli Esposti, C. Giardin\`a, S. Graffi}\\
%{\large M. Degli Esposti, C. Giardin\`a, S. Graffi}\\
{\small\it
Dipartimento di Matematica \\ Universit\`a di Bologna,\\
Piazza di Porta S. Donato 5, 40127 Bologna, Italy \\
}
\end{center}
\date{}
%
%\begin{abstract}
%\end{abstract}
%
 \abstract{The Random Orthogonal Model (ROM) of
Marinari-Parisi-Ritort \cite{MPR1,MPR2} is a model of statistical
mechanics where the couplings among the spins are defined by a
matrix chosen randomly within the orthogonal ensemble. It
reproduces the most relevant properties of the Parisi solution of
the Sherrington-Kirckpatrick  model. Here we compute the energy
distribution, and work out an extimate for the two-point
correlation function. Moreover, we show exponential increase of
the number of metastable states also for non zero magnetic field.
}

\section{Introduction: review of the model and outlook}

Random (symmetric) matrices out of a given ensemble can be taken
as interaction matrices for Ising spin models. The most famous
example is the Sherrington-Kirckpatrick (SK) model of spin
glasses, where the elements are i.i.d. Gaussian variables with
properly normalized variance. Aim of this paper is
 to discuss a very specific example of these spin glass models,
which also share some interesting connections with number theory,
and show how random matrix theory could be useful to investigate
 its
properties.

 For the sake of simplicity, let us start with a very
concrete
 question: let $N\geq 1$ be a positive integer and denote
$\SN$ the
 space of all possible configurations of $N$ spin
variables

$$
\SN=\{\s=(\s_1,\ldots,\s_N)\,\,,\,\,\s_j=\pm
1\},\quad\vert\SN\vert=2^N.
$$
Given $k=1,\ldots,N-1$, denote  $C_k$ the correlation function:
$$
C_k(\s)\,=\,\sum_{j=1}^{N}\s_j\s_{j+k},\quad\mbox{where }
j+k:=(j+k-1\mod N) +1,
$$
and define the Hamiltonian function
$$
H(\s)\,=\,\frac{1}{N-1}\sum_{k=1}^{N-1} C_k^2
$$
For each N the ground state of the Hamiltonian $H$ can be looked at
as the binary
 sequence with  lowest autocorrelation and finding it
will have
 some relevant practical applications in efficient
communication
 (see \cite{Be} and references in \cite{MPR1}).

It is remarkable that no  concrete procedure for reproducing the
ground state for generic $N$ is known, but  {\it ad hoc}
constructions based on number theory exist for very specific
values of $N$:  if $N$ is  prime number with $N=3\mod 4$, then the
sequence of the Legendre
symbols\footnote{$\left(\frac{j}{N}\right)=1$, if $j=x^2\,\mod N$
and $-1$ otherwise.} ($\s_N=1$)
$$
\s_j:=\left(\frac{j}{N}\right)=j^{\frac{1}{2}(N-1)}\mod N\,,\quad
j=1,\ldots,N-1
$$
 gives the ground state of the system
\cite{DGGI,MPR1}.

Through the use of the discrete Fourier transform, it is not
difficult to see \cite{MPR1,PP} that the previous problem is in
fact equivalent to finding the ground state for the so called {\it
Sine model}, which represents our starting point:
$$
H(\s)\,=\,-\frac{1}{2}\sum_{i,j=1}^N J_{ij}\s_i\s_j
.
$$
Here $J$ is the following $N\times N$ real symmetric orthogonal
matrix with almost full connectivity:

$$
J_{ij}\,=\,\frac{2}{\sqrt{1+2N}}\,\sin\left(\frac{2\pi\,ij}{2N
+1}\right),\quad i,j=1,\ldots,N
$$
\vskip.3cm\noindent

 Here again, if $2N+1$ is prime and $N$ odd,
the Legendre symbols $\s_j=j^N\mod 2N+1$ give the ground state of
the system for these very specific values of $N$.

A natural approach  is to extend the study of the ground state to
the more general thermo-dynamical behavior of the model in terms
of the inverse temperature $\beta=\frac{1}{T}$. As usual, the two
basic objects are: $$ \mbox{\it the partition function}\quad
Z_J(\beta)\,:=\,\sum_{\sigma\in\SN} e^{-\beta H(\sigma)}, $$ and
 {\it the free energy density (at the thermodynamical
limit)} $$ f_J(\beta)\,=\,\lim_{N\to\infty}-\frac{1}{\beta N} \log
Z_J(\beta) $$\vskip0.3cm\noindent

It is important to remark now that even if {\it there is no
randomness in the system}, the ground state of the model {\it
looks like} an output of a random number generator and the numeric
of its thermo-dynamical properties resemble the one of disordered
systems. This observation was in fact the starting point of an
approach developed in \cite{MPR1,MPR2,PP} where this  model is
seen as a particular realization of a disordered model where the
coupling matrix is chosen at random out of a suitable set of
matrices:
\begin{definition} The {\it Random Orthogonal Model} (ROM) with magnetic field $h\geq0$ is the
disordered system with energy \be\label{hamilton}
H_J(\s)\,=\,-\frac{1}{2}\sum_{ij}J_{ij}\s_j\s_i\,+
\, h\sum_{j}\s_j, \ee
where
the  coupling matrix $J$ is chosen randomly in the set of
{\it orthogonal symmetric matrices}:\footnote{In the ROM model
generic matrices have non zero diagonal elements. Often these
terms will be set to zero and orthogonality will be reconstructed
in the large $N$ limit.}
$$
J\,=\,ODO^{-1},
$$
Here $O$ is a generic orthogonal matrix and $D$ is diagonal with
entries $\pm 1$. The numbers $\pm 1$ are the eigenvalues of $J$.

The natural probability measure $\mu$ on this set is
the product of
 the canonical Haar measure on the orthogonal group
by the
 discrete measure on the diagonal terms.
\end{definition}
\vskip .5cm We will use the notation $\brac{\cdot}$ to denote the
average with respect the measure $\mu$. In particular, we are
interested in the {\it quenched}  (i.e., the
average is performed after taking the logarithm) free energy density:

\be\label{quenchedf}
\brac{f_J(\beta)}\,=\,-\lim_{N\to\infty}\frac{1}{\beta N}
\brac{\log Z_J(\beta)} \ee

 Average over the ROM disorder is performed by the following
 fundamental formula, which has been obtained by adapting the
 results in \cite{IZ} (see also \cite{BG}) valid for the unitary
case to the orthogonal one \cite{MPR2}.
 For any  $N\times N$
symmetric matrix $A$:
 \begin{eqnarray} \brac{\exp \left\{ tr
\left(\frac{JA}{2}\right)\right\}} &=& \exp \left\{ N tr \left(
G\left(\frac{A}{N}\right)\right)\right\} +R_N(A), \nonumber\\
&\cong&
 \exp \left\{ N \sum_{j=1}^N
G\left(\lambda_j\right)\right\}\label{main} \end{eqnarray} where
$R_N\to 0$ in the thermodynamical limit $N\to\infty$, the
$\lambda_j$'s are the (real) eigenvalues of $\frac{1}{N}\cdot A$
and $G(x)$ is given by $$ G(x) = \frac{1}{4}\left[ \sqrt{1+4x^2}
-\ln \left(\frac{1+\sqrt{1+4x^2}}{2}\right)-1\right] $$

The same formula is exact for the SK model, i.e Gaussian
independent symmetric couplings,  with $$ G_{SK}(x)=\frac{x^2}{4}
$$ Note that $G(x) = G_{SK}(x) + o(x)$. For example, up to the
10-th order
$$
G(x)\,=\,\frac{x^2}{4} - \frac{x^4}{8} + \frac{x^6}{6} -
\frac{5\,x^8}{16} +
  \frac{7\,x^{10}}{10} + O(x^{11})
  $$

   The ROM model has been chosen in a such a way that, at least for
 not too small temperature, the {\it deterministic} Sine model and
 the one with quenched disorder share a common behavior. More
 precisely, the couplings are always of order $N^{-\frac{1}{2}}$;
 the diagrams contributing to the thermodynamical limit of the high
 temperature expansion for the free energy density have
the same topology and they can all be
 expressed in terms of positive powers of the trace of the
 couplings. By construction, the high temperature expansion of the
 free energy density $f_J(\beta)$ in powers of $\beta$ is then
 independent of the particular choice of the symmetric orthogonal
 matrix $J$ and it does coincide with the annealed average w.r.t.
 $\mu$. In particular \cite{PP}:
 $$
 -\beta \brac{f_J(\beta)}\,=\, \log 2 + G(\beta).
 $$

Besides SK and in general the large class of $p$-spin models,
where couplings have a gaussian distribution, the ROM model
provide another interesting class of disordered mean-field spin
glass. This model has received considerable interest in recent
years, especially in the contest of structural glass transition.
Indeed it can been seen as the random version of a wide class of
models (for example the fully frustrated Ising model on a
hypercube or the above mentioned sine model) which despite having
a non-random Hamiltonian display a strong glassy behavior
\cite{BM,MPR1,DGGI}. This model has been studied in the framework
of replica theory \cite{MPR2}, where it was shown that replica
symmetry is broken and there are many equilibrium states available
to the system. Mean field (TAP) equations have been derived for
this model by resumming the high temperature expansion and the
average number of solutions of these equations has been studied in
ref. \cite{PP}.

It is a  well established fact that the observed properties of
mean-field spin glass models are due to the large number of {\it
metastable states} the system possesses. Despite being not fully
justified from a mathematical point of view, the Parisi scheme of
breaking replica symmetry furnishes a clear picture of equilibrium
statistical properties: states with similar macroscopic behavior
have vastly different spin configurations, and have large
relaxation times for transition between them. As a consequence,
the ground state is accessible only on very long time scales. It
is worth mentioning
 that rigorous results validating the Parisi solution
have been
 accumulating in recent times.

For example, Guerra and Toninelli\cite{G1} have proved the existence
of thermodynamical limit, i.e. the existence of
the limit for quenched
 average of the free energy (eq.
\ref{quenchedf}). See also
 \cite{CDGG} where the result has been
extended to general
 correlated gaussian random energy models.
Finally, more recently
 \cite{G2}, Guerra  showed that the
Parisi Ansatz represents
 at least a lower bound for the quenched
average of the free
 energy.

However there is not yet an unambiguous way to identify
those metastable
 states which are relevant for thermodynamics in the
infinite volume
 limit. At zero temperature the metastable states can
be defined as
 the states {\it locally stable} to single spin
flips (definition recalled in Section 3 below) and
the  calculations
are relatively straightforward. Complete analysis of
 the typical
energy of metastable states and of the effects of the
 external
field have been undertaken both for the SK model
 \cite{TE,Ro,D} and
for general $p$-spin model \cite{OF}. The zero
 temperature dynamics
for the deterministic {\it Sine model} has
 been instead studied in
\cite{DGGI}.

At non-zero temperature the identification is less obvious and
most studies \cite{BrM,Ri} rely on the counting of the number of
solutions to the celebrated TAP equations \cite{TAP}. According to
the general belief, one can associate to each metastable state a
solution of the TAP equation, but the inverse is not true: a TAP
solution corresponds to a metastable state only if it is separated
from other solution by a barrier whose height diverges with the
volume.

It appears, however, that even the calculations at zero
temperature in a presence of external field have not been carried
out. One expects, in analogy with SK model, the existence of an AT
line \cite{AT} indicating the onset of replica-symmetry breaking.
In this paper we study at length the effects of the magnetic field
on the structure of local optima of the energy landscape. We are
able to use these results to shed further light on the nature of
the AT instability at zero temperature.

 In the next Section \ref{statistics} we study the statistics of energy levels over the
whole configuration space. We compute energy distribution of a
generic spin configuration and the pair correlations for a given
couple of spin configurations with a fixed overlap. In Section
\ref{metastable} we analyze metastable states at zero temperature,
also in a presence of an external field.

\section{Statistics of energy levels}
\label{statistics}

We start by analyzing  the statistical features of the landscape
generated by the energy function (\ref{hamilton}). In this section
we will always consider zero magnetic field $h=0$. Let us begin
with the energy distribution for a single fixed configuration.

\subsection{Distribution of energy}

Let $\s = (\s_1,\s_2,\ldots,\s_N)$ denote a given configuration
with energy $H_J(\s)$.  The probability $P_\s(E)$ is then given
by: $$ P_\s(E) \,:=\, \< \delta(E - H(J,\s))\> $$

\noindent Due to gauge invariance, the probability $P_\s(E)$ does
not depend on the spin configuration $\s$ and it will be denoted
just by $P(E)$, in fact: $H(J,\s) = H(J',\s')$ and $P(J) = P(J')$
where $J_{ij}' = J_{ij}\s_i\s_i'\s_j\s_j'$.

\noindent Introducing the integral representation for $\delta$
function

$$ \delta(x - x_0) = \frac{1}{2\pi i}\int_{-i \infty}^{+i \infty}
dk \; e^{k \; (x-x_0)} $$
 we get

$$ P(E) = \frac{1}{2\pi i}\int_{-i \infty}^{+i \infty} dk \; e^{kE}
\; \<e^{\frac{1}{2} \sum_{i,j =1}^{N}k \; J_{ij}\s_i\s_j}\> $$

\noindent and we can apply formula (\ref{main}) to average over
disorder considering the matrix $A_{ij} = k\s_i\s_j$.

It is easy to prove that  $A$ admits only one non-zero, simple
eigenvalue $\lambda=kN$, so that
$$P(E) = \frac{1}{2\pi i}\int_{-i
\infty}^{+i \infty} dk \; \exp \left[N \left(\frac{kE}{N} +
G(k)\right)\right] $$

\noindent In the large-$N$ limit the integral can be evaluated
using the saddle-point method. Clearly,  the equation
$$
\frac{E}{N} + G'(k)\,=\,\frac{E}{N} + \frac{k}{1 + {\sqrt{1 +
4\,k^2}}}=0
$$
admits the solution $\bar{k}= \frac{2\,E\,N}{4\,E^2 - N^2}$.

 This gives:

\begin{eqnarray}
 P_{ROM}(E) &\sim& \exp\left[N\left(\frac{\bar{k}E}{N} +
G(\bar{k})\right)\right]\label{rom1}\\
 &=& \left(1-\left(\frac{2E}{N}\right)^2\right) ^{N/4}\nonumber\\
 &\sim&\
\exp\left[-\frac{E^2}{N}-2\frac{E^4}{N^3}-\frac{16}{3}\frac{E^6}{N^5}+\cdots\right]\nonumber\\
 \end{eqnarray}

\noindent apart for an unimportant constant, not predicted by the
saddle-point. As a comparison, in the case of SK model one finds
exactly the gaussian distribution: $$ P_{SK}(E) \sim \exp
\left(-\frac{E^2}{N}\right) $$

\begin{figure}[ht]
\includegraphics[width=4in,angle=270]{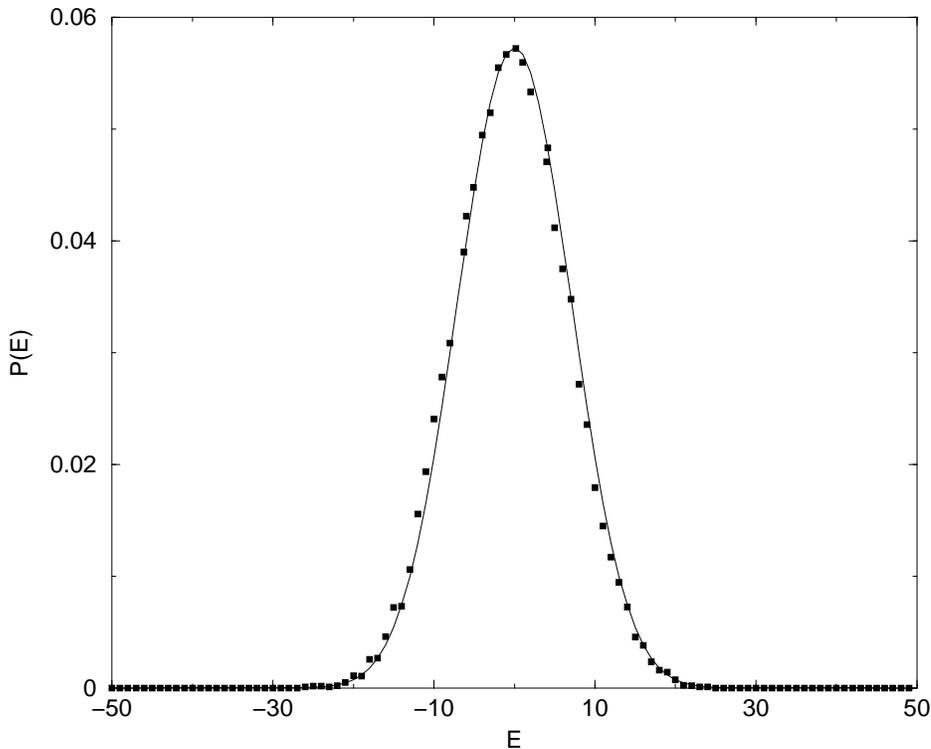}
\caption{\baselineskip=13pt {\small { Probability distribution
function $P_{ROM}(E)$ for ROM model (full curve). Simulation for a
$N=100$ ROM (data points). For a fixed spin configuration, $10^6$
realizations of disorder were generated.}}} \protect\label{fig1}
\end{figure}

To check the validity of formula (\ref{main}) which has been used
to average over disorder, we computed $P_{ROM}(E)$ for a relative
small ROM (N=100) numerically. For a given spin configuration,
random disorder realizations $J=ODO^{-1}$ were generated by using
an orthogonal
 matrix $O$  obtained from a
gaussian matrix by applying Gram-Schmidt orthogonalization
algorithm and coin tossing for the diagonal $D$. The resulting
distribution of energies was binned and is shown as the data
points in Fig. (\ref{fig1}).

\noindent
%The agreement between analytical calculation and this
%simulation is very good, so we consider this a fair test of
%formula (\ref{main}).

 As it should be, the support of $P_{ROM}(E)$ is almost all
 in the interval $[-N/2,N/2]$. Indeed, the orthogonality of
$J$ imposes simple bound on the energy of any spin configuration:
the lower bound $-N/2$ (resp. upper bound $N/2$) is reached if and
only if $\s$ is an eigenvector of $J$ relative to eigenvalue $+1$
(resp. $-1$).

\subsection{Two-point Energy Correlation}

We consider now the probability $P_{\s,\tau}(E_1,E_2)$ that two
configurations $\s,\tau\in\SN$  have energies $E_1$ and $E_2$
respectively. Gauge invariance implies that this probability can
only depend on the overlap between the two configurations:

$$ q(\s,\tau) = \frac{1}{N}\sum_{i=1}^N \s_i\tau_i $$

\noindent Proceeding as before, we get: \bea P_{\s,\tau}(E_1,E_2)
& = & \< \delta(E_1 - H(J,\s)) \;
                    \delta(E_2 - H(J,\tau))\>
           \nonumber\\
     & = & \frac{1}{(2\pi i)^2}\int_{-i \infty}^{+i \infty} dk_1 \;
           \int_{-i \infty}^{+i \infty} dk_2 \;
            \exp (k_1 E_1+k_2 E_2) \nonumber\\
     &   &  \<\exp \left(\frac{1}{2} \sum_{i,j =1}^{N} J_{ij}
            (k_1 \s_i\s_j + k_2 \tau_i\tau_j)\right)\>
\eea

\vspace{0.5cm} \noindent Consider now the matrix $A_{ij} = k_1
\s_i\s_j + k_2 \tau_i\tau_j$ which has two non-zero simple
eigenvalues

$$ \lambda_{\pm} = \frac{N}{2}\left[(k_1+k_2) \pm
                 \sqrt{(k_1-k_2)^2 + 4k_1k_2q^2} \right]
$$

\noindent Applying formula (\ref{main}) we obtain: \bea
P_{\s,\tau}(E_1,E_2) & = & \frac{1}{(2\pi i)^2}\int_{-i
\infty}^{+i \infty} dk_1 \;
                 \int_{-i \infty}^{+i \infty} dk_2 \nonumber \\
           &   & \exp \left[ N \left( \frac{k_1 E_1}{N} + \frac{k_2 E_2}{N}
                 + G(\lambda_{+}/N) + G(\lambda_{-}/N) \right) \right
                 ]\nonumber
\eea

The saddle-point method yields the equations:
$$
\frac{E_j}{N} +\frac{1}{N}\,
G'(\frac{\lambda_+}{N})\,\frac{\partial\lambda_+}{\partial k_j}
\,+\,\frac{1}{N}\,
G'(\frac{\lambda_-}{N})\,\frac{\partial\lambda_-}{\partial
k_j}\,=\,0,\,\,\qquad j=1,2
$$

For the  SK model, one immediately find:
$$
\frac{E_1}{N}+\frac{1}{2}(k_1+k_2
q^2)=0,\quad\frac{E_2}{N}+\frac{1}{2}(k_2+k_1 q^2)=0
$$
with solutions:
$$
k_1=\frac{2(E_1-E_2 q^2)}{N(-1+q^4)},\quad k_2=\frac{2(E_2-E_1
q^2)}{N(-1+q^4)}.
$$
This yields the well known \cite{De} ($ \s,\t\in\SN$ fixed, with
overlap $q$):
\begin{eqnarray} P_{SK}(E_1,E_2)  &=& \left(\frac{\sqrt{1-q^4}}{N\pi}\right)\exp \left
[-\frac{(E_1+E_2)^2}{2N(1+q^2)}\right]
                       \exp \left
                       [-\frac{(E_1-E_2)^2}{2N(1-q^2)}\right]\nonumber\\
                       &=&
P_{SK}\left(\frac{E_1+E_2}{\sqrt{2(1+q^2)}}\right)\cdot
P_{SK}\left(\frac{E_1-E_2}{\sqrt{2(1-q^2)}}\right).
\end{eqnarray}
For asymptotically uncorrelated configurations $q=0$, one clearly
get a product measure, whereas  one recover complete degeneracy
when $q=1$: \be \label{sk1}P_{SK}(E_1,E_2)\,=\, P_{SK}(E_1)\cdot
P_{SK}(E_2),\quad q=0,
 \ee
and \be \label{sk2}P_{SK}(E_1,E_2)\,=\,
P_{SK}(E_1)\cdot\delta\left(E_2-E_1\right),\quad q=1.
 \ee

 In generale, one has
 $$
 \int_{-\infty}^{+\infty}\,\int_{-\infty}^{+\infty} E_1E_2\,
 dP_{SK}(E_1,E_2)\,=\,\frac{Nq^2}{2}.
 $$

For the ROM model, it is immediate to see that the analog of
(\ref{sk1}) and (\ref{sk2}) hold true with  the single energy
distribution $P_{ROM}(E)$ given by (\ref{rom1}). For generic value
of $0<q<1$, a first crude estimate is achieved by using the
stationary points of the gaussian approximation and
$G(x)=\frac{x^2}{4}-\frac{x^4}{8}$ to evaluate the exponent. This
yields:
$$
P_{ROM}(E_1,E_2)\sim P_{SK}(E_1,E_2)\cdot Exp[-\Phi_q(E_1,E_2)],
$$
where
\begin{eqnarray}
\Phi_q(E_1,E_2)\,&:=&\,-2\frac{\, -8\,E_1^3\,E_2\,q^4 -
      8\,E_1\,E_2^3\,q^4 +
      E_1^4\,\left( 1 + 2\,q^2 - q^4 \right)   }{N^3\,
    {\left( -1 + q^2 \right) }^2\,{\left( 1 + q^2 \right)
    }^4}\nonumber\\
    &&+\frac{
      {E_2}^4\,\left( 1 + 2\,q^2 - q^4 \right)  +
      2\,{E_1}^2\,{E_2}^2\,q^2\,
       \left( 2 - q^2 + 4\,q^4 + q^6 \right)}{N^3\,
    {\left( -1 + q^2 \right) }^2\,{\left( 1 + q^2 \right)
    }^4}\nonumber
    \end{eqnarray}

    Further corrections can be now calculated, but we do not known
    a systematic way of doing it at all orders.

\section{Zero temperature metastable states}
\label{metastable}

Metastable states at zero temperature are defined as the
configurations whose energy can not be decreased by reversing any
of the spins \cite{DGGI}. Since the energy change $\Delta E_i$
involved in flipping the spin at site $i$ is given by $$ \Delta
E_i = 2 \left(\sum_{j}J_{ij}\s_i \s_j + h\s_i\right) $$ the
constraint a configuration $\s$ must satisfy in order to be
metastable is $$\sum_{j}J_{ij}\s_i \s_j + h\s_i > 0 \quad\quad
\forall i=1,\ldots, N$$

\noindent The average number of metastable configurations $\<{\cal
N}(e,h)\>$ with a given energy density $e=E/N$ is then

\bea \<{\cal N}(e,h)\>
         & = & \< \sum_{\{\s\}}
               \prod_{i=1}^N
               \left[ \int_{0}^{\infty} d\lambda_i \;
               \delta \left(\lambda_i - \sum_{j}J_{ij}\s_i\s_j - h\s_i\right)
                \right] \nonumber \\
         &   &  \delta\left( Ne + \frac{1}{2} \sum_{i,j}J_{ij}\s_i\s_j
               + h \sum_{i} \s_i \right)  \>
\eea

\noindent One should really calculate the average value of the
logarithm of the number of metastable states, this being the
extensive quantity, and hence introduce replicas; indeed, as
pointed out in \cite{BrM}, the introduction of a uniform magnetic
field should introduce strong correlations between the metastable
states. However, we shall proceed with the direct calculation of
$\<{\cal N}(e,h)\>$ as it suffices to bring out the most relevant
features of the problem.

\noindent Introducing integral representations for $\delta$
functions we have \bea \<{\cal N}(e,h)\>
         & = & \sum_{\{\s\}}
               \int_{-i\infty}^{+i\infty} \frac{dz}{2\pi i} \;
               e^{z Ne} e^{z h \sum_{i} \s_i} \nonumber \\
         &   & \prod_{i=1}^N
               \left[ \int_{0}^{\infty} d\lambda_i \;
                \int_{-i \infty}^{+i \infty} \frac{dk_i}{2\pi i} \;
                \right]
                e^{\sum_{i} k_i(h\s_i-\lambda_i)} \;
               \< e^{ \sum_{i,j} J_{ij}
                (\frac{z}{2}\s_i\s_j  + k_i\s_i\s_j}) \> \nonumber
\eea

\noindent

\noindent To apply the formula (\ref{main}) for averaging over
disorder we define the matrix $A_{ij} = \left( \frac{z}{2} + k_i
\right) \s_i\s_j + \left( \frac{z}{2} + k_j \right)\s_j\s_i$. The
non-zero eigenvalues of $A_{ij}$ are easily calculated and read

$$\mu_{\pm} = \sum_i \left( \frac{z}{2} + k_i \right) \pm \sqrt{N
\sum_i \left( \frac{z}{2} + k_i \right)^2} $$

\noindent so that we obtain:

\bea \<{\cal N}(e,h)\> & = &  \sum_{\{\s\}}
\int_{-i\infty}^{+i\infty} \frac{dz}{2\pi i}\;
               e^{z Ne} e^{z h \sum_{i} \s_i}\prod_{i=1}^N
         \left[ \int_{0}^{\infty} d\lambda_i \;
                \int_{-i \infty}^{+i \infty}\frac{dk_i}{2\pi i} \;
                \right] \;
                e^{\sum_{i} k_i(h\s_i-\lambda_i)}  \nonumber\\
          &   & \exp\left\{N\left[
                G\left(\frac{1}{N}
                \sum_i \left( \frac{z}{2} + k_i \right) +  \sqrt{\frac{1}{N}
                \sum_i \left( \frac{z}{2} + k_i \right)^2}  \right)
                \right.\right. + \nonumber\\
          &   & \left.\left.
                G\left(\frac{1}{N}
                \sum_i \left( \frac{z}{2} + k_i \right) -  \sqrt{\frac{1}{N}
                \sum_i \left( \frac{z}{2} + k_i \right)^2}  \right)
                \right]\right\}
\eea

\noindent Performing now the trace over spin configuration,
defining

$$v = \frac{1}{N}\sum_i \left( \frac{z}{2} + k_i \right)
\quad\quad\quad w = \frac{1}{N}\sum_i \left( \frac{z}{2} + k_i
\right)^2 $$

\noindent and imposing the constraints via two Lagrange
multipliers, we have

\bea \<{\cal N}(e,h)\> & = & \frac{1}{(2\pi i)^3}
                \int_{-i \infty}^{+i \infty}dz \;
                \int_{-i \infty}^{+i \infty}dv \;
                \int_{-i \infty}^{+i \infty}dw \;
                \int_{-i \infty}^{+i \infty}dx \;
                \int_{-i \infty}^{+i \infty}dy \;
                \nonumber \\
          &   & \exp\left\{
                N [ ze + \frac{zx}{2} + \frac{yz^2}{4}]
                \right\} \nonumber \\
          &   & \exp\left\{
                N [ -xv -yw + G(v + \sqrt{w})+G(v - \sqrt{w})]
                \right\} \nonumber \\
          &   & \prod_{i=1}^N
                \left[ \int_{0}^{\infty} d\lambda_i \;
                \int_{-i \infty}^{+i \infty}\frac{dk_i}{\pi i} \;
                e^{y k_i^2 + k_i(x-\lambda_i + yz)}
                \cosh{(h(z + k_i))}\right]
\eea

\noindent The integrals over the $k_i$ are now gaussian \bea
\<{\cal N}(e,h)\>  & = & \frac{1}{(2\pi i)^3}
                \int_{-i \infty}^{+i \infty}dz \;
                \int_{-i \infty}^{+i \infty}dv \;
                \int_{-i \infty}^{+i \infty}dw \;
                \int_{-i \infty}^{+i \infty}dx \;
                \int_{-i \infty}^{+i \infty}dy \;
                \nonumber \\
          &   & \exp\left\{
                N [ ze + \frac{zx}{2} + \frac{yz^2}{4}]
                \right\} \nonumber \\
          &   & \exp\left\{
                N [ -xv -yw + G(v + \sqrt{w})+G(v - \sqrt{w})]
                \right\} \\
          &   & \prod_{i=1}^N
                \left[ \int_{0}^{\infty} d\lambda_i \;
                \frac{1}{2\sqrt{\pi y}} \left(
                e^{hz} e^{-\frac{(x+yz-\lambda_i+h)^2}{4y}} +
                e^{-hz}e^{-\frac{(x+yz-\lambda_i-h)^2}{4y}}
                \right)\right]\nonumber
\eea

\noindent and the integrals over the $\lambda_i$ can be performed
in terms of the complementary error function

$$ \mbox{erfc(x)} = \frac{2}{\sqrt{\pi}}
                 \int_{x}^{\infty}  e^{-t^2}\,dt
$$
\noindent so that we find: \bea \label{saddle} \<{\cal N}(e,h)\>
& = & \frac{1}{(2\pi i)^3}
                \int_{-i \infty}^{+i \infty}dz \;
                \int_{-i \infty}^{+i \infty}dv \;
                \int_{-i \infty}^{+i \infty}dw \;
                \int_{-i \infty}^{+i \infty}dx \;
                \int_{-i \infty}^{+i \infty}dy \;
                \nonumber \\
          &   & \exp\left\{
                N [ ze + \frac{zx}{2} + \frac{yz^2}{4}]
                \right\} \nonumber \\
          &   & \exp \left\{N \left[ -xv -yw + G(v + \sqrt{w})+G(v - \sqrt{w})
                \right.\right.\\
          &   & + \ln \left(\frac{1}{2}
                \left(e^{hz}\mbox{erfc}\left(-\frac{x+yz+h}{2\sqrt{y}}\right) +
                    e^{-hz}\mbox{erfc}\left(-\frac{x+yz-h}{2\sqrt{y}}\right)
                 \right) \right) \left.\left.\right]\right\} \nonumber
\eea

\noindent As usual the calculation is concluded by carrying out a
saddle-point integration. The r.h.s. of Eq. (\ref{saddle}) is to
be extremized with respect to the five variables $z,v,w,x,y$.

\subsection{Total number of metastable states}

\begin{figure}[ht]
\includegraphics[width=3in]{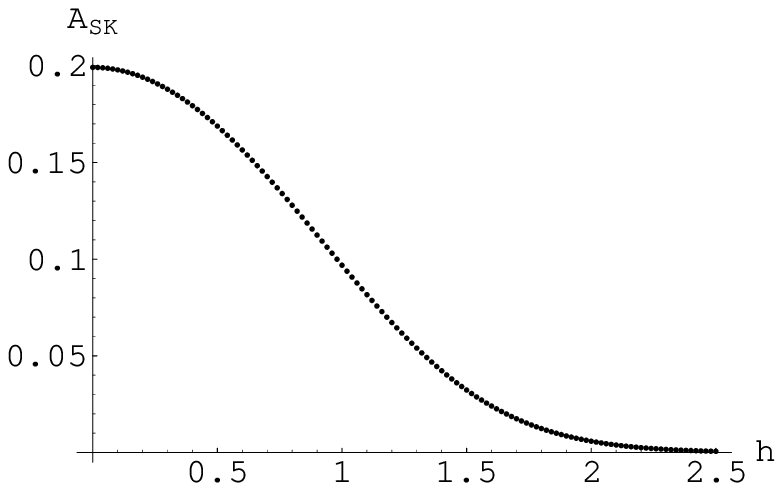}
\caption{\baselineskip=13pt {\small { $A_{SK}(h)$ }}}
\protect\label{metask}
\end{figure}

Here we study the  total number of metastable states ${\< \cal
N}(h) \>$ (irrespectively of the energy) as a function of the
field. Writing
 \be\label{Ah}
\log \< {\cal N}(h)\> = A(h)N + B(h), \ee $A(h)$, in the
thermodynamical limit $(N\rightarrow \infty)$, can be calculated
by setting $z=0$ in Eq. (\ref{saddle}), which becomes:

\bea \<{\cal N}(h)\>  & = & \frac{1}{(2\pi i)^2}
                \int_{-i \infty}^{+i \infty}dv \;
                \int_{-i \infty}^{+i \infty}dw \;
                \int_{-i \infty}^{+i \infty}dx \;
                \int_{-i \infty}^{+i \infty}dy \;
                \nonumber \\
          &   & \exp \left\{N \left[ -xv -yw + G(v + \sqrt{w})+G(v - \sqrt{w})
                \right.\right.\nonumber \\
          &   & + \ln \left(\frac{1}{2}
                \left(\mbox{erfc}(-\frac{x+h}{2\sqrt{y}}) +
                 \mbox{erfc}(-\frac{x-h}{2\sqrt{y}})\right)
                \right)
                \left.\left.\right]\right\}\label{saddle1}
\eea In the case of the SK model one recover the well now
one-variable saddle-point equation \cite{D}:
$$
x\,=\,\frac{\exp[-x^2/2] \cosh(h x)}{\int_{-x}^{\infty}
\exp[-t^2/2]\cosh(h t)\,dt}
$$
If $x_c$ is the solution to the previous equation:
$$
A_{SK}(h)\,=\,\log(2)\,-\,\frac{1}{2}(x_c^2+h^2)\,+\,\log\left(\frac{1}{(2\pi)^{1/2}}\int_{-x_c}^{\infty}
\exp[-t^2/2]\cosh(h t)\,dt\right),
$$
in particular $A_{SK}(0)\sim 0.199$, whereas for large $h$ one has
$x\sim\left(\frac{2}{\pi}\right)^{1/2}\, e^{-h^2/2}$ and
consequently $A_{SK}\sim\frac{1}{\pi}e^{-h^2}$ (see
Fig.(\ref{metask})).

\begin{figure}[!h]
\includegraphics[width=2.1in,angle=270]{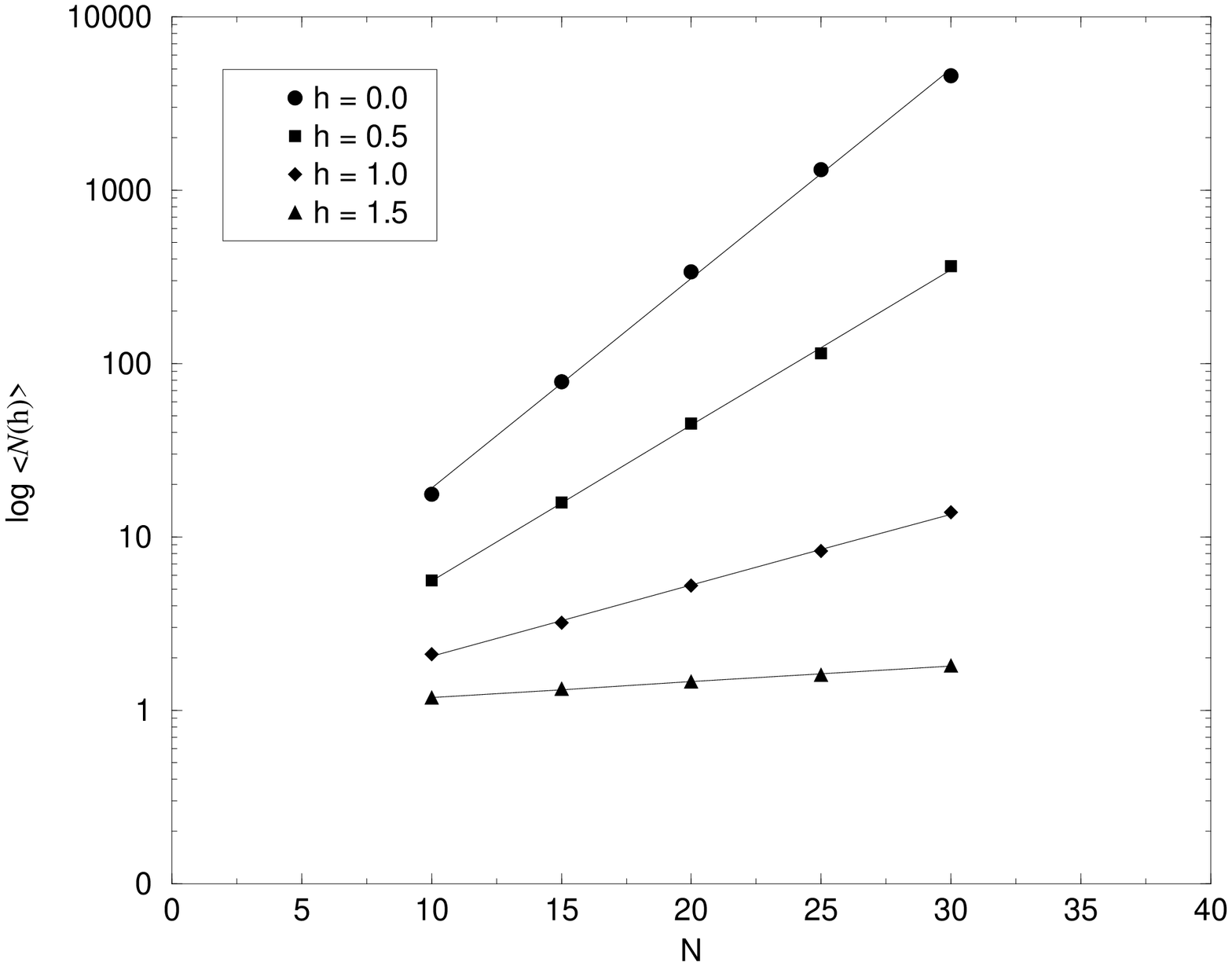}
\includegraphics[width=2.1in,angle=270]{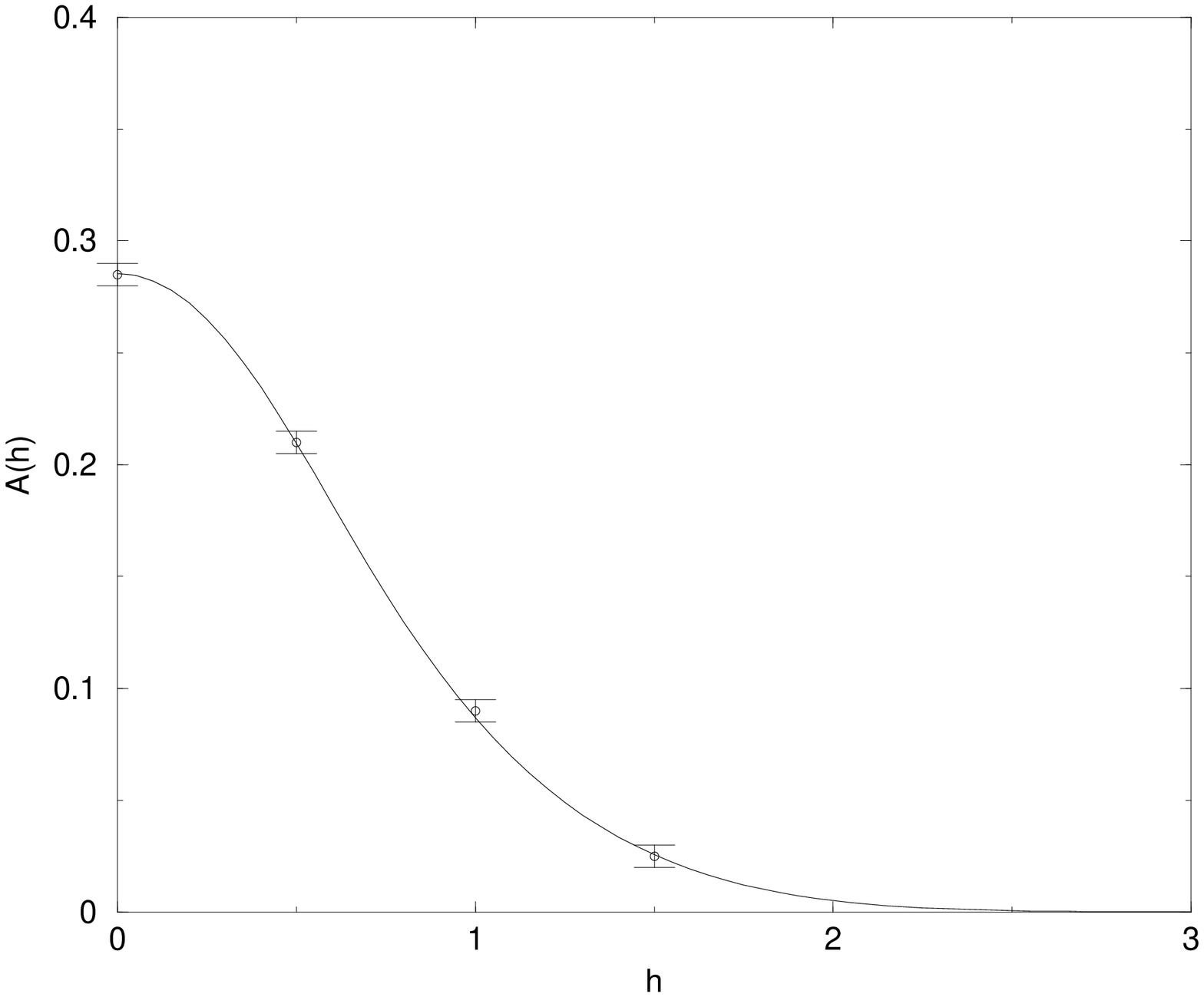}
\caption{\baselineskip=13pt {\small { Numbers of metastable
states. (Left): Their dependence on system size $N$ at different
magnetic field, see legend. (Right): Data points shows the field
dependence of $A_{ROM}(h)$ obtained from the fits, while the full
curve indicates the analytical results in the thermodynamical
limit. }}} \protect\label{fig2}
\end{figure}

We now turn to the ROM model. We first perform a numerical
investigation by doing an exhaustive enumeration of spin
configurations and keeping track of metastable states. The
system-size dependence of $\log \< {\cal N}(h)\>$ is plotted for
different values of $h$ in Fig. (\ref{fig2}, left). The data are
fitted to formula (\ref{Ah}), ignoring possible finite size
corrections. The resulting $A_{ROM}(h)$ are showed in Fig.
(\ref{fig2}, right) as data points.

\noindent Moreover, the saddle point equations corresponding to
(\ref{saddle1}) were solved numerically, and the result is shown
by the solid curve in (Fig. (\ref{fig2}, right)). The agreement
between theory and simulations is very good in spite of the fact
that we used admittedly small systems ($N<30$).

\noindent As one would expect, metastable states disappear as the
magnetic field is increased, since it introduces a tendency
towards ferromagnetic behavior. Most of the processes are the
confluence of a metastable state to another with a larger drop of
free-energy.

\begin{figure}[!h]
\includegraphics[width=3in]{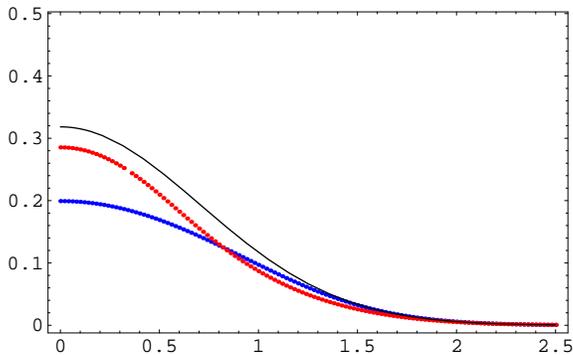}
\caption{\baselineskip=13pt {\small { $A_{SK}(h)$ (bottom
blue),$A_{ROM}(h)$ (middle red ), $\frac{1}{\pi}e^{-h^2}$ (top)
.}}} \protect\label{romfig1}
\end{figure}

\begin{figure}[!h]
\includegraphics[width=3in]{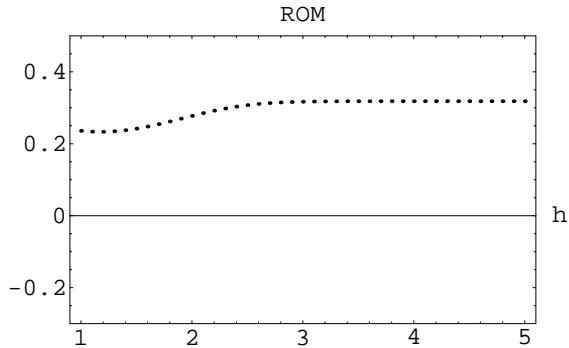}
\caption{\baselineskip=13pt {\small { Plot of $e^{h^2}\cdot
A_{ROM}(h)$ for values of the magnetic field $h$ between $1$ and
$5$.}}} \protect\label{romfig2}
\end{figure}

Note that we have $A_{ROM}(0)\sim 0.285$, while the asymptotic
behavior for large magnetic field $h$ does coincide with the
gaussian case (see Fig.(\ref{romfig1},Fig.(\ref{romfig2} )).

This indicates that $A_{ROM}(h)$ still remain non-zero for
arbitrarily large $h$ and hence for any finite value of the
external magnetic field the number of metastable states grow
exponentially with the system size $N$. As pointed out by \cite{D}
for the SK model, this result is in agreement with the observation
that the AT instability occurs for all finite $h$ at zero
temperature.

\thanks{{\bf Acknowledgements:} This work has been partially
supported by the European Commission under the Research Training
Network MAQC,
 n° HPRN-CT-2000-00103 of the IHP Programme.}

\end{document}